\documentclass[english, 11pt]{article}
\usepackage[latin9]{inputenc}
\usepackage{babel}
\usepackage{color}
\usepackage{geometry}
\title{Carter Constant and Angular Momentum}
\author{Sajal Mukherjee and K. Rajesh  Nayak,  \\
IISER Kolkata, Mohanpur, India..}
\begin{document}
\maketitle
\begin{abstract}
We investigate the Carter-like constant in the case of a particle  moving  in a  non-relativistic  dipolar potential.  This  special case is a missing link  between Carter constant in stationary axially symmetric spacetimes such as Kerr solution and its possible  Newtonian counterpart.  We use this system to  carry over the definition of  angular momentum from the Newtonian mechanics  to the  relativistic  stationary axially symmetric  spacetimes. %\textcolor{blue}{A new approach to write the total angular momentum for axially symmetric stationary spacetime  is also  introduced.} 
\end{abstract}
%\pacs{04.20.Cv, 04.70.Bw}
\section{Introduction}
The Carter constant is one of the non-trivial integrals of motion
for a  particle moving in a stationary axially symmetric  spacetimes~(SASS)
such as Kerr solution~\cite{carter68}. Though there have been several studies on the physical interpretation, a self-consistence understanding of this integral of motion  has been a problem. In the case of simpler spherically  symmetric solutions, the Carter constant reduces to the square of the angular momentum~($L^2=\vec{L}\cdot\vec{L}$)~\cite{Defici}. One of the obstacles in understanding the Carter constant is the complexity of the  SASS, because of the  physical effects such as frame-dragging~\cite{LT1}. The phenomenon frame-dragging  is essentially a pure general relativistic effects  that  do not have any  Newtonian analogy.  For a simpler understanding and to investigate non-trivial properties  of the Carter constant, it would be good  to have a solution to Einstein equations which is axially symmetric and static.  However, attemptst to find such solutions,  allowing Carter constant  are  not very successful~\cite{MSWill}.   Another  approach is to look for a Newtonian system  that would give rise to a non-trivial  Carter-like constant. Recently, Will~\cite{will2009} has given Carter-like constant in the Newtonian dynamics. Here, we take a similar approach and show that  Carter-like constant exists for a charged particle moving in a field of an electric dipole. Perhaps, this case can be used for better understanding the physics of   Carter constant in general.  We also look at the Carter constant in the case of relativistic rotating frame.  In this standard example, the Carter constant reduces to $L^2$. 

The angular momentum vector is a purely Newtonian  concept, and it plays an 
important roll in  understanding physics of a rotating system. However,   it does not have  any covariant definition,   and this has been one of the  problems in the general theory of relativity.   The  angular momentum vector, because of the interrelation between it's  components, only  two independent scalar quantities can  be constructed.    Conventionally  they  are  the azimuthal component of the angular momentum, $L_z$  and  square of the angular momentum,  $L^2$.  Until  a valid covariant definition of angular momentum is established, one may use $L_z$ and $L^2$ for understanding the role of angular momentum in the relativistic rotating systems. In the case spherical symmetry,  both  $L_z$ and $L^2$ are conserved, and it is very easy to identify them. However,  in the case SASS,  $L_z$ is conserved and in general $L^2$ need  not be conserved. In such case,  there is no straightforward way to identify $L^2$.  It is here  we use the Carter constant  with its Newtonian counterpart to define  $L^2$ in SASS. 
One of the interesting special case would be, a  SASS  that  allows $L^2$ as a conserved quantity. We show that  only  spherically symmetric and  flat spacetime allow such possibility. 

The central point of this article is around   a theorem given by Carter~\cite{CART1973},  it states,  for a Hamiltonian the form: 
\begin{equation}
\mathcal{H}=\frac{1}{2}\frac{H_{r}+H_{\theta}}{U_{r}+U_{\theta}}\,,\label{eq:cham}
\end{equation}
where, $U_{r}$, $U_{\theta}$ are functions of single variable $r$,
$\theta$ respectively. The function $H_{r}$ is function of $r$
and canonical momenta other than $p_{\theta}$, while $H_{\theta}$
is function of $\theta$ and canonical momenta other than $p_{r}$,
then 
\begin{equation}
\mathcal{K}=\frac{U_{r}H_{\theta}-U_{\theta}H_{r}}{U_{r}+U_{\theta}}\,,\label{eq:cart}
\end{equation}
is a constant of motion. This integral  of motion $\mathcal{K}$ was named after Carter. 
This theorem  is a generalised version of St\"ackel theorem, which put the constraint on the form of  potentials giving rise to  conserved quantity through the Hamilton-Jacobi theory~\cite{Stackel}.  

In this article, we  apply the Carter's theorem  to three examples. In the first case, for a charged particle moving in the  field of an electric dipole. This  system is a well understood Newtonian axially symmetric system. In section-\ref{sec:acc_frame}, we apply it to a particle moving in a rotating frame. Finally in the section-\ref{sec:sass},  we apply it to the case of 
SASS and Kerr spacetime as a special case. We show that the Newtonian definition of $L^2$ can be 
carried over to SASS.    Finally,  we close the  article  with a brief  concluding remark.

%%%%%%%%%%%%%%%%%%%%%%%%%%%%%%%%%%%%%%%%%
\section{Carter-St\"ackel constant for a dipolar field \label{sec:dipole}}
We apply the Carter theorem for the motion of a  charged particle in
an   axially symmetric electrostatic  field. The Most suitable coordinates  to study axially symmetric system is the spherical polar coordinates~$\left\{ r,\:\theta,\:\phi\right\} $.
 The non-relativistic Hamiltonian  in the spherical polar coordinate takes the form:
\begin{equation}
\mathcal{H}=\frac{1}{2mr^{2}}\left\{ \left(rp_{r}\right)^{2}+\left(p_{\theta}\right)^{2}+\frac{1}{\sin^{2}\theta}\left(p_{\phi}\right)^{2}+2mr^{2}\Phi\left(r,\theta\right)\right\}\,. \label{eq:Hamilt}
\end{equation}
We use the following choice of decomposition for $\mathcal{H}$, in
terms of $\{H_{r},\, H_{\theta},\: U_{r},\: U_{\theta}\}$: 
\begin{eqnarray}
U_{r}=r^2\,, &  \ & U_{\theta} =0 \,,\nonumber \\
H_{r}=r^{2}p_{r}^{2} & \ &H_{\theta}= \left(p_{\theta}\right)^{2}+\frac{1}{\sin^{2}\theta}\left(p_{\phi}\right)^{2}+2r^{2}\Phi\left(r,\theta\right)\,.
\label{eq:Decom} 
\end{eqnarray}
Form separability point of view, we required that $\Phi\left(r,\theta\right)=\frac{1}{r^{2}}g\left(\theta\right)$.
Where  $g\left(\theta\right)$ can be any
arbitrary function of $\theta$~\cite{rana}. This form of potential is  known as  St\"ackel potential~\cite{rana}. 
From the physics point of view, the choice of potential is:  
\begin{equation}
\qquad\Phi\left(r,\theta\right)=\frac{a}{r^{2}}\cos\theta\,.\label{eq:dipole}
\end{equation}
In the above,  $a$ is a constant, and  the potential is  due to a point   dipole.   The Newtonian gravity itself does not allow dipolar field;   we can consider an example from classical electrodynamics,  a point charge particle in the field of point electric  dipole.  The  corresponding conserved  quantity is: 
\begin{equation}
\mathcal{K}_{dipole}=H_{\theta}=\left[p_{\theta}^{2}+\frac{1}{\sin^{2}\theta}p_{\phi}^{2}\right]+2 a \cos\theta \,\, = L^2 +2 a \cos\theta \,.\label{eq:Cart_dipole}
\end{equation}
We may note that $p_{\phi}=L_{z}$, the azimuthal component of the angular momentum. When the field is spherically symmetric, {\it i.e.} if $a=0$, the Carter constant reduces to $L^2$. The dipolar field gives the lowest non-trivial contribution to the Carter-St\"ackel constant. It is the case where conservation of $L^2$  is not valid. Since, in the Newtonian mechanics, angular momentum is defined  unambiguously, we can identify the two terms, the first one as $L^2$ and the second one as  effective potential for motion along $\theta$.
                                                                                                                                                                                                                                                       
The motion of a charged particle in electric dipole field has been well
studies starting from Turner and Fox~\cite{TAF1968}.  However, the existence of an additional integral of motion was first explicitly emphasised  by Gutierres-Lopez, Castellanos-Moreno and Rosas-Burgos~\cite{GCR2008}, without dwelling on  its close association with the Carter constant. They have given an in-depth analysis  of the effect of this constant on the motion of charge particle. In particular, we would like to draw attention to the maximum angle subtended by the charged particle while moving on a sphere ($r=$constant ) for a given value of Carter constant. Similar behaviour is also  exhibited by a particle moving in the Kerr spacetime~\cite{Wilkins}.  

The precession is one of the  consequences of non-conservation of angular momentum.  In our case, the  precession equation can  be obtained  by taking the  derivative of Eq~(\ref{eq:Cart_dipole}), resulting in:
\begin{equation}
\vec{L}\cdot \dot{\vec{L}}=-\beta q \sin\theta\, \dot{\theta}
\label{eq:lldot}
\end{equation}
This relation gives the precession of angular momentum for a particle orbiting around a dipole, implying that the motion of the particle is not confined to a single plane.  

%%%%%%%%%%%%%%%%%%%%%%%%%%%%%%%%%%%%%%%%%
\section{Rotating Frame\label{sec:acc_frame}}
In this section,  we apply the Carter's theorem  to a particle moving  in a  relativistic rotating frame. The spacetime metric with a rotation along the $z-axis$ with constant angular velocity $\omega$ can be written as:
\begin{equation}
ds^2=-\left(1- \omega^2 r^2  \sin^2\theta\right)\,dt^2 - 2 \omega r^2 \sin^2\theta \, dt d\phi + r^2 \sin^2 \theta \, d\phi^2 +dr^2 + r^2 \, d\theta^2\,.
\label{eq:metrot} 
\end{equation}
The spacetime metric components are functions  of the  coordinates and will 
have at least one off-diagonal component.  However, the absence of  gravity is  reflected as vanishing   Riemann curvature, {\it i.e.},  $R_{abcd}\equiv0$.
In this case, the Hamiltonian for a  particle can be written as: 
\begin{equation}
\mathcal{H}=\frac{1}{r^2}\left[ -r^2 p_t^2 - 2\omega r^2 p_t p_\phi + \left(\frac{1}{\sin^2\theta} - \omega^2 r^2 \right) p_\phi^2 + r^2 p_r^2 +p_\theta^2 \right]\,.
\label{eq:hamrot} 
\end{equation}
It is straight forward to apply  the Carters theorem,  resulting in: 
\begin{equation}
\mathcal{K}_{rotate}=\left[p_\theta^2 + \frac{1}{\sin^2\theta}p_\phi^2\right]\ =\ L^2\,.
\end{equation}
 In this case, both  $L_z$  and  $L^2$ are conserved. Irrespective of the
 rotation of the frame itself  here the Carter constant reduces to  $L^2$.
We can easily extend the  dipole potential given in the section-\ref{sec:dipole} to a 
rotating frame. This case will not change the overall  scenario, as the rotation does not affect potential. However, the rotating systems in the general relativity introduce 
more complexity that we discuss in the next section.

%%%%%%%%%%%%%%%%%%%%%%%%%%%%%%%%%%%%%%%%%%
\section{Stationary Axially Symmetric  Spacetimes\label{sec:sass}}
In this section, we apply the Carter's theorem to  the case of   a particle moving in  SASS. Here, we closely follow the  Carter's work~\cite{CART1973}.  It has been shown that, for a spacetime to allow Carter constant, it is sufficient that the spacetime metric takes the following form\cite{CART1973}: 
\begin{eqnarray}
ds^2\, = \, \Sigma\left\{ \frac{dr^2}{\Delta_r} + \frac{d\mu^2}{\Delta_\mu} \right\}+\frac{1}{\Sigma}\left\{\Delta_\mu \left[ C_r dt - Z_r d\phi \right]^2 - \Delta_r \left[ C_\mu dt - Z_\mu d\phi \right]^2\right\}\,.
\label{eq:assst}
\end{eqnarray}
In the above,  $ \Sigma=\left(C_{\mu} Z_r - C_r Z_{\mu} \right)$,  $C_r$ and $C_\mu$ are constants. The parameters   $Z_\mu$ and $\Delta_\mu$ are functions of single variable $\mu=\cos\theta$.  While, the parameters  $Z_r$ and $\Delta_r$ are functions of $r$ only~\cite{CART1973}.  These functions can be obtained by solving the Einstein equations. For example, in the case of  the Kerr solution~\cite{KERR}, these parameters are given by: 
\begin{eqnarray}
 C_\mu=1\,, & \ & C_r=a\,, \nonumber \\
 Z_r=r^2+a^2\,, & \ & Z_\mu=a\left(1-\mu^2\right)\,, \nonumber \\
 \Delta_r=r^2-2Mr+a^2\,, & \ & \Delta_\mu=1-\mu^2\,.
\label{eq:kerrst}
\end{eqnarray}
The metric given in Eq.~(\ref{eq:assst}), allows  a time Killing  vector field $\xi^a$ and a rotational Killing vector $\eta^a$. The   corresponding  conserved quantities are, energy    $E=p_a\,\xi^a$ and the azimuthal component of  angular momentum $L_z=p_a\,\eta^a$.

The Newman-Penrose formalism is  very useful in studying the Carter constant. For the metric given in Eq. (\ref{eq:assst}),  components of the  complex null tetrad are given by \cite{np}:
\begin{eqnarray}
l^a &= &  \frac{1}{\Delta_r}\left( Z_r, \Delta_r, 0, C_r\right)\,, \nonumber \\
n^a &=& \frac{1}{2\Sigma} \left(Z_r, -\Delta_r, 0, C_r \right)\,, \nonumber \\
m^a &=& \frac{1}{\sqrt{2\Sigma}}\left( \frac{i Z_\mu}{\sqrt{\Delta_\mu}},
0, \sqrt{\Delta_\mu}, \frac{i C_\mu}{\sqrt{\Delta_\mu}} \right)\,.
\label{eq:NPfr}
\end{eqnarray}
The null vectors $l^a$ and $n^a$ are principal null directions~\cite{CHANDRA,KSM}.
The spacetime metric in terms of the null tetrad is given by:
\begin{equation}
g^{ab}=  -\left( l^a n^b+l^b n^a - m^b\bar{m}^a-m^a \bar{m}^b\right)\,.
\label{eq:nullmetric}
\end{equation}
In the above, $\bar{m}^a$ is the complex conjugate of $m^a$.
In this case, the Hamiltonian $\mathcal{H}$, takes the form:  
\begin{eqnarray}
\mathcal{H}=\frac{1}{\Sigma}\left\{  \Delta_\mu  p_\mu^2 +\Delta_r p_r^2  + \Delta_\mu^{-1}\left[C_\mu p_\phi +Z_\mu p_t \right]^2  - \Delta_r^{-1} \left[ C_r p_\phi + Z_r p_t  \right]^2\right\}
\label{eq:asshelton}
\end{eqnarray}  
It is straightforward to compute the Carter constant using Eq. (\ref{eq:cham}) and Eq. (\ref{eq:cart})~\cite{CART1973}. After some simplification, we get:
\begin{equation}
\mathcal{K}_{sass} = \left[ \Delta_\mu p_\mu^2 + \frac{C_\mu^2}{\Delta_\mu} p_\phi^2 \right] + \left[ \frac{Z_{\mu}^2}{\Delta_\mu} E^2 -2 \frac{C_\mu Z_\mu}{ \Delta_\mu } L_z E -\frac12 C_r Z_\mu  \right] \,.
\label{eq:Ksass}
\end{equation}
Here, we use  normalising condition $\mathcal{H}=g^{ab}p_ap_b=-\frac12$. Interestingly, this equation looks very similar to the dipole case as given in Eq.(\ref{eq:Cart_dipole}). In the above equation we have,  first part which is quadratic in momenta of angular motion, {\it i.e.} along  $\mu$ and $\phi$. In analogy with Eq.~(\ref{eq:Cart_dipole}), we may  recognise it as  $L^2$. The second part,  depend only on $\mu$ and constants of motion such as $E$ and $L_z$.  It  is associated  with the  effective potential for motion along  $\mu$.   From this close similarity, we  define  $L^2$ in the case of axially symmetric stationary spacetimes as,
\begin{equation}
L^2 = \left[ \Delta_\mu p_\mu^2 + \frac{C_\mu^2}{\Delta_\mu} p_\phi^2 \right] \,.
\label{eq:L2}
\end{equation}
It is just the Newtonian definition carried over to the relativistic case, which  is valid  for SASS with the metric given by the Eq.(\ref{eq:asshelton}). The above equation gives the definition of $L^2$ in terms of generalised momenta, $p_\theta$ and $p_\phi$. It should be noted that  form of  $p_\theta$ and $p_\phi$  might be   different   in each case.  Here, we would like to remind that,  in the case of SASS, the $p_a$ in terms of $\dot{x}^a$ are given by:
\begin{eqnarray}
\arraycolsep=1.4pt\def\arraystretch{1.8}
\begin{array}{ lclcl}
p_t &=&g_{tt} \dot{t} + g_{t\phi} \dot{\phi} &=& \frac{\Delta_\mu C^2_r-\Delta_r C^2_\mu}{\Sigma}\dot{t} \,+\frac{\Delta_r C_\mu Z_\mu -\Delta_\mu C_r Z_r}{\Sigma}\dot{\phi}\,,\\
p_r & = & g_{rr}\,\dot{r}&=& \frac{\Sigma}{\Delta_r}\dot{r}\,,\\
p_\mu & = & g_{\mu\mu}\,\dot{\mu}&=&\frac{\Sigma}{\Delta_\mu} \dot{\mu}, \\
p_\phi  & = & g_{t\phi} \dot{t} + g_{\phi\phi} \dot{\phi}&=&\,\frac{\Delta_r C_\mu Z_\mu -\Delta_\mu C_r Z_r}{\Sigma}\dot{t} \,+\frac{\Delta_\mu Z_r^2-\Delta_r Z^2_\mu}{\Sigma}\dot{\phi}.
\end{array}
\label{eq:pa}
\end{eqnarray}
Though   $L^2$ takes the simpler form similar to Newtonian case, the complexity of  relativistic rotation is still maintained.  
In the case of Kerr spacetime, from  Eq.~(\ref{eq:kerrst}) and Eq.~(\ref{eq:Ksass}), we get
this well known expression for Carter constant:
\begin{equation}
\mathcal{K}_{kerr} = \left[ p_\theta^2 + \frac{1}{\sin^2\theta} \, p_\phi^2 \right] +  a^2 \sin^2\theta E^2  \, -2aELz \, -\frac{a^2\sin^2\theta}{2}\, .
\label{eq:Kerr}
\end{equation} 
It should be noted that, $L^2= \left[ p_\theta^2 + \frac{1}{\sin^2\theta} \, p_\phi^2 \right]$, comes from Eq.~(\ref{eq:L2})  and is  not as a result of setting  parameter $a\rightarrow0$.

In the past, there have been attempts to define angular momentum through the Floyd tensor\cite{Faridi, Ram}. However, we take a simpler approach, by  defining the  quantities $L_z$ and $L^2$.  As has been mentioned earlier, these are the two physically important scalar quantities connected  with angular momentum vector.
In general, $L^2$ is not conserved,  if the separability conditions are satisfied as per Carter's theorem, we have additional conserved quantity $\mathcal{K}_{sass}$.    Physically it indicates that,   in general,  the motion of a free particle in SASS is not confined to a plane.  Here, 
we have obtained this  definition of  $L^2$ from analogy and intuition. In the next section, we arrive at the same result with a more formal approach. 

%%%%%%%%%%%%%%%%%%%%%%%%%%%%%%%%%%%%%%%%%%%%%
\subsection{Relation with Killing Tensor}
%%%%%%%%%%%%%%%%%%%%%%%%%%%%%%%%%%%%%%%%%%%%%
In this section, we approach the definition of $L^2$ in SASS using the Killing tensor. 
The Carter constant can also be obtained from the Killing tensor \cite{walker}. The Killing tensor in terms of the principal  null vectors  for SASS is given by,
\begin{equation}
K^{ab}=\Sigma \left( l^a n^b + l^b n^a \right) + C_\mu Z_r \,g^{ab}\,.
\label{eq:Kten1}
\end{equation}
For a particle moving along geodesic  with four-momentum $p^a$, $\mathcal{K}=K^{ab} p_a p_b $ is 
conserved and can be shown to be Carter constant given by Eq. (\ref{eq:Ksass}).  We show that the Killing tensor  given in Eq.~(\ref{eq:Kten1}), can be casted  in a  following alternative form:
\begin{equation}
K^{ab}=\Sigma \left(m^a \bar{m}^b + m^b \bar{m}^a \right) + C_r Z_\mu\, g^{ab}\,.
\label{eq:Kten1}
\end{equation} 
It is straightforward to show that the $K^{ab}$ indeed a Killing tensor. After some 
simplification it can be shown that the conserved quantity $\mathcal{K}=K^{ab} p_a p_b $ is same 
as the one given in Eq. (\ref{eq:Ksass}).
 Because of the normalisation condition $g^{ab}p_a p_b=-\frac{1}{2}$,  the  last term is Eq.~(\ref{eq:Kten1}) independent of $p_a$ and   can not contribute to   $L^2$. In the definition of $L^2$, we consider only the  first term,  which is:
 \begin{equation}
L^{ab}=\Sigma \left(m^a \bar{m}^b + m^b \bar{m}^a \right)\,.
\label{eq:Lab}
\end{equation}
 We project this onto  two-surface orthogonal to the surface formed by $\chi^a$  and
$\gamma^a$.  Where,
\begin{equation}
 \chi^a=\xi^a-\frac{\xi^p \eta_p}{\eta^q \eta_q} \eta^a \ \ \ \mathrm{ and } \  \ \
\gamma^a=\left(0,\,1,\,0,\,0\,\right)\, .
\end{equation}
 The vector field $\gamma^a$  is  along the radial direction.  In SASS,  the time like-Killing vector  $\xi^a$ and  rotational Killing vector $\eta^a$  are surface forming~\cite{GSV}. However, they are not orthogonal, {\it i.e.},  
 $\xi^a \eta_a \neq 0$.  Besides,  if spacetime satisfy the condition of  orthogonal transitivity \cite{carter1969},  $r\,-\theta\,$ surfaces at every point in the spacetime are   orthogonal to $\xi-\eta$ surfaces. As  $\xi^a$ and $\eta^a$ are not orthogonal, we span the same surface with the vector field $\chi$ and $\eta$.  The vector field $\chi^a$ is not a Killing  vector field and referred as quasi-Killing vector field~\cite{GSV}.  It   has several interesting properties,  it is hypersurface orthogonal and can be  taken as frame closest
to the Newtonian global rest frame in SASS \cite{ ANW2}.  Furthermore, they are also 
referred as zero angular momentum observers~(~ZAMO~) because observers following along $\chi$ experiences no precession or has zero angular momentum~\cite{BRD}. By using this projection, we ensure that no total angular momentum is lost during the projection operation and remain close  to a possible Newtonian description. The projection operator  onto a $\theta\,-\phi$ surface is given by,
\begin{equation}
h^{ab}=g^{ab}-\frac{1}{\chi^p \chi_p} \chi^a \chi^b -\frac{1}{\gamma^q \gamma_q} \gamma^a \gamma^b\,.
\label{eq:hab}
\end{equation}  
With these, the $L^2$ can be defined as:
\begin{equation}
L^2=L^{ab} h_a^{\ k}h_b^{\ l} \ p_k\, p_l\,,
\label{eq:L2def}
\end{equation}
where $L^{ab}$ is given by Eq.~(\ref{eq:Lab}).
It is easy to show this takes the same  form given in the Eq. (\ref{eq:L2}).

%%%%%%%%%%%%%%%%%%%%%%%%%%%%%%%%%%%%%%%%%%%%%
\subsection{Special cases}
%%%%%%%%%%%%%%%%%%%%%%%%%%%%%%%%%%%%%%%%%%%%%
Here, we  look for   a special case of SASS   in which $L^2$ is conversed. 
We start with  the  case $Z_\mu=0$,  for which $\mathcal{K}_{sass}$ reduces to just $L^2$.  These are  a possible  set  of  axially symmetric system that conserves $L^2$.  For this case, the  spacetime metric takes the following form: 
\begin{equation}
ds^2 = \frac{-1}{C_\mu Z_r} (\Delta_r C^2_\mu -\Delta_\mu C_r^2 ) dt^2 - 2 \frac{\Delta_\mu C_r}{C_\mu} dt d\phi + \frac{\Delta_\mu Z_r}{C_\mu} d\phi^2 + \frac{C_\mu Z_r}{\Delta_r} dr^2 + \frac{C_\mu Z_r}{\Delta_\mu} d\mu^2
\label{eq:Zmuzeromet}
\end{equation}
We look for source free solutions to the  Einstein's equation satisfied by the above metric. The significant constraint comes from  $G^{(\mu)} _{(\mu)}$ component of the  Einstein tensor.  The equation $G^{(\mu)} _{(\mu)}=0$  implies: 
\begin{equation}
\left(\frac{d\,Z_r}{dr \ } \right)^2 \left(C_\mu^2 \Delta_r - C_r^2 \Delta_\mu \right) + 2 C^2_\mu Z_r \left(Z_r  \frac{d^2\,\Delta_r}{dr^2 \ } - \frac{d\, Z_r}{dr \ }  \frac{d\,\Delta_r}{dr \ } \right) = 0
\end{equation}
$\Delta_\mu$  is the only parameter depending  on the  variable $\mu$.  There are two ways obtaining  a self-consistent solution:
\begin{enumerate}
\item $C_r=0$: In this case,  it is  straightforward to show that spacetime is spherically  symmetric,  resulting in the known result $\mathcal{K}=L^2$.
\item $C_r\neq0$: This can be shown to result in the condition that the Riemann tensor $R_{abcd}=0$,  {\it i.e.}, the space-time is flat. This case may include cases such as rotating frame has given earlier or any other transformation resulting in the metric of the form given in Eq.(\ref{eq:Zmuzeromet}).
\end{enumerate}
Leading to a completely consistent  cases of $L^2$ conservation for the 
source free solutions~\cite{MSWill, MTW}.

%%%%%%%%%%%%%%%%%%%%%%%%%%%%%%%%%%%%%%%%%%
\section{Discussions}
A particle moving in a dipolar field serves as a good Newtonian system for  understanding the Carter constant.  We have shown that the form of the Newtonian Carter-St\"ackel constant has  interestingly close similarity with its counterpart in fully relativistic SASS such as Kerr solution. This Analogy allows us carry over the definition of $L^2$ from Newtonian mechanics to SASS. We would like to emphasise that only form of the metric is important for this formalism, it does not depend on the nature of the source or asymptotic conditions.
Because it is possible to have Carter-like constant for SASS which are not asymptotical flat~\cite{Ram}. 
More formally, the $L^2$ can be defined using Killing tensor and its projection onto a space orthogonal to the space formed by the ZAMO  and radial vector.  In the case of source free solutions, we also have shown that only spacetimes  admitting   conservation of $L^2$  are: spherically symmetric system and the spacetime with zero  Riemann curvature. 
%This not neccesserilty means there no, axially symmetric static system
 %In an axially symmetric system, In general $L^2$ is not conserved.  If the spacetime metric components satisfy the Carter separability condition,  $\mathcal{K}$ is conserved.   The Carter constant , $\mathcal{K}$,  will have two terms. First term can be identified as   $L^2$  and  the second term, $V_\mu$,   depends on  only $\mu(=\cos\theta)$. The latter  term is  connected with  effective potential for the motion along $\theta$.
%\newpage

\section{Acknowledgments}
We wish to thank the Visiting Associateship programme of 
Inter-University Centre for Astronomy and Astrophysics~(IUCAA), Pune.   A
part of this work was carried out during the visit to IUCAA under this programme. 

\end{document}